# Effects of pressure on the local atomic structure of $CaWO_4$ and $YLiF_4$: Mechanism of the scheelite-to-wolframite and scheelite-to-fergusonite transitions

D. Errandonea[1,2], F.J. Manjón[3,†], M. Somayazulu[2], and D. Häusermann[2]

[1] *Departamento de Física Aplicada-ICMUV, Universitat de València, Edificio de Investigación, c/Dr. Moliner 50, 46100 Burjassot (Valencia), Spain*

[2] *HPCAT, Carnegie Institution of Washington, Advanced Photon Source, Building 434E, Argonne National Laboratory, 9700 South Cass Ave., Argonne, IL 60439, U.S.A.*

[3] *Departamento de Física Aplicada, Universitat Politècnica de València, Pl. Ferrandiz i Carbonell 2, 03801 Alcoy (Alicante), Spain*

**Abstract:** The pressure response of the scheelite phase of $CaWO_4$ ($YLiF_4$) and the occurrence of the pressure induced scheelite-to-wolframite (M-fergusonite) transition are reviewed and discussed. It is shown that the change of the axial parameters under compression is related with the different pressure dependence of the W-O (Li-F) and Ca-O (Y-F) interatomic bonds. Phase transition mechanisms for both compounds are proposed. Furthermore, a systematic study of the phase transition in 16 different scheelite $ABX_4$ compounds indicates that the transition pressure increases as the packing ratio of the anionic $BX_4$ units around the A cations increases.

## 1. INTRODUCTION

Many ABX$_4$ compounds, like calcium tungstate (CaWO$_4$) and yttrium lithium fluoride (YLiF$_4$), crystallize in the tetragonal scheelite structure (SG: *I4$_1$/a*, No. 88, Z=4) **[1, 2]** at ambient conditions. The strong interest in the structural stability of scheelite compounds under compression is evident by the numerous experimental studies on the pressure effects on their phase behavior **[3 - 14]**. In particular, it has been demonstrated recently that CaWO$_4$ transforms under compression from the scheelite structure to the monoclinic wolframite structure (SG: *P2/c*, No. 13, Z=2) **[1, 2]** at 11 ± 1 GPa **[3, 4]**. On the other hand, YLiF$_4$ transforms under compression from the scheelite structure to the monoclinic M-fergusonite structure (SG: *C2/c*, No. 15, Z=4) **[1, 2]** also at 11 ± 1 GPa **[5, 6]**. In both compounds, the reversibility to the initial scheelite structure after decreasing pressure has been shown.

From the cationic point of view, the scheelite structure consists of two intercalated diamond lattices: one for A cations and another for B cations (see **Fig. 1**), where the A-A distances are equal to B-B distances. In the scheelite structure, A cations, calcium (Ca) and yttrium (Y), are coordinated by eight X anions, oxygen (O) or fluorine (F), thus forming AX$_8$ polyhedral units. On the other hand, B cations, tungsten (W) and lithium (Li), are coordinated by four X anions forming relatively isolated BX$_4$ tetrahedral units **[7]**. In the cation coordination notation for ABX$_4$ compounds ([cation A coordination – cation B coordination]), scheelites have cation coordination [8 – 4]. **Fig. 1** shows a detail of the scheelite structure with the AX$_8$ and BX$_4$ polyhedra.

The study of the pressure effects on the local atomic structure can be a powerful tool to understand the transformation mechanisms of pressure-driven transitions. While a systematic analysis of the effects of pressure on the local atomic structure of YLiF$_4$ has already been performed **[5]**, the same analysis in CaWO$_4$ has not been performed yet. In this work, we report and discuss the pressure response of the local structure of W (Li) ions in CaWO$_4$ (YLiF$_4$) on the light of recently reported high-pressure x-ray diffraction data **[3, 5]** and other high-pressure techniques. The aim of discussing the effects of pressure in the local structure of both compounds is to understand more precisely the occurrence of the scheelite-to-monoclinic transitions, and particularly, the scheelite-to-wolframite and scheelite-to-fergusonite transitions. From the characterization of the similarities and differences of the pressure response of the local structure of CaWO$_4$ and YLiF$_4$ possible transformation mechanisms for both transitions are identified.

---

† Corresponding author: fjmanjon@fis.upv.es, Tel.:(+34) 96 652 8442, Fax:(+34) 96 652 8409

**2. EXPERIMENTAL BACKGROUND**

The lattice parameters and bond distances here presented for CaWO$_4$ were obtained from energy-dispersive x-ray powder diffraction (EDXD) patterns measured at the X-17C beamline at the National Synchrotron Light Source (NSLS) using a diamond-anvil-cell (DAC) at a diffraction angle 2θ = 13°. As CaWO$_4$ is soft (bulk modulus, B$_0$ = 77 **[3]**), this material was used as its own quasi-hydrostatic pressure medium. A detailed description of these experiments was given in Ref. **[3]**. There, we reported the occurrence of the scheelite-to-wolframite transition of CaWO$_4$ at 11 GPa and the amorphization of it at 40 GPa, but we did not discuss the pressure effects on the local structure of the scheelite phase of CaWO$_4$. In the present paper, we report a detail analysis of this issue, by comparing the pressure response of the local structure of CaWO$_4$ and YLiF$_4$, in order to understand better the pressure behavior of the structure in scheelite-type ABX$_4$ compounds. We show in **Fig. 2** an x-ray diffraction pattern of $CaWO_4$ measured at 2 GPa. The spectrum is plotted together with the difference between the measured data and the calculated profile with the aim of illustrating the quality of the structural refinements used to extract the lattice parameters and bond lengths of $CaWO_4$ here presented. In order to obtain the lattice parameters from the experimental data the Le Bail extraction technique [15] available in the GSAS programme [16] was employed. For every analyzed pressure, we obtained good agreement between the refined profiles and the experimental diffraction patterns, as illustrated in **Fig. 2**, and a low value for the residual for the intensities, $R(F) < 0.15$ (for 52 reflections). The bond distances for $CaWO_4$ were calculated after performing the structural refinements using the POWDERCELL programme package [17]. The analogous data on $YLiF_4$, used for the comparative analysis of the pressure effects on the scheelite compounds $CaWO_4$ and $YLiF_4$, were obtained from Ref. [5]. This recent work reported data obtained from angle-dispersive powder diffraction experiments performed at the ID9 beamline at the European Synchrotron Radiation Facility using a monochromatic beam ($\lambda = 0.4203$ Å) and a DAC with methanol-ethanol as pressure medium.

## 3. RESULTS AND DISCUSSION

### 3.1 Pressure effects on the local atomic structure

In order to know the microscopic mechanisms governing the scheelite-to-monoclinic phase transitions in $CaWO_4$ and $YLiF_4$ we have analyzed the pressure dependence of the lattice parameters and bond distances in these two compounds. **Fig. 3** shows the pressure dependence of the lattice parameters for the scheelite phase of CaWO$_4$ and YLiF$_4$. Both compounds show a clearly anisotropic character, the compressibility of the *c* axis being larger in CaWO$_4$, and the compressibility of the *a* axis being larger in YLiF$_4$. This behavior is reflected in **Fig. 4**, which shows that the *c/a* ratio in both compounds evolves in a different way under pressure, being *c* more compressible than *a* in CaWO$_4$ while the contrary is true for YLiF$_4$. The *c/a* axial ratio decreases under compression from 2.17 at ambient conditions (1 bar) **[8]** to 2.136 at 11.3 GPa in CaWO$_4$ **[3]**, but it increases from 2.08 at 1 bar to 2.12 at 11 GPa in YLiF$_4$ **[5]**. This different behavior of the *c/a* ratio under pressure in CaWO$_4$ and YLiF$_4$ was previously noted by the different linear compressibilities of the lattice parameters measured in these two compounds **[18]**.

In order to better understand the different anisotropic behavior of both scheelites under pressure, it is very useful to describe them in terms of the pressure response of the AX$_8$ and BX$_4$ polyhedra. With this aim, the pressure dependence of the W-O distances inside the BX$_4$ tetrahedra and the Ca-O distances inside AX$_8$ polyhedra are plotted for CaWO$_4$ in **Fig. 5**. The small pressure dependence of the W-O distance, as compared to that of the Li-F distance (see Fig. 4 in Ref. **[5]**), indicates that WO$_4$ tetrahedra are rigid and isolated structural elements that undergo little change with pressure up to 11 GPa, unlike LiF$_4$ tetrahedra, which are more compressible in the same pressure range. On the other hand, Ca-O (Y-F) bond compression is significantly greater (smaller) than that of W-O (Li-F) bonds. These differences in the compressibilities are the cause of the decrease (increase) of the *c/a* axial ratio in CaWO$_4$ (YLiF$_4$).

It is well known that application of pressure reduces the interatomic distances and the atomic sizes, being the large anions more compressible than the small cations **[19, 20]**. Therefore, the effect of pressure is twofold:

i) with increasing pressure the decrease of the interatomic distances and of cation sizes leads to an increase of the cation-cation repulsive forces **[7, 21]**; and

ii) the reduction of anion sizes leads to an increase of the packing efficiency of anions in the cationic sublattice.

According to Sleight **[7]**, the increase of the cation-cation repulsion forces leads to a decrease of the *c/a* ratio tending to 2 in tetragonal $ABX_4$ compounds. This *c/a* value corresponds to that of the ideal structure for equal near-neighbors cation-cation distances and consequently to equal cation coordination. On the other hand, the increase of the anion packing efficiency leads to an increase of the *c/a* ratio and consequently to different cation coordination numbers.

Based upon these considerations, we think that the effect of pressure on the phase transitions depends greatly on which of the two above mechanisms predominate with the increase of pressure: cation-cation repulsion or anion packing efficiency. In this sense, it must be noted that the axial ratio of the scheelite structure of $YLiF_4$ at atmospheric pressure is closer to *c/a* = 2 than that of $CaWO_4$. Moreover, with increasing pressure this latter compound tends to the ideal structure for equal cation coordination while the former separates from it. On this basis, it can be concluded that the high-pressure phase transition of scheelites and the cation coordination of the high-pressure phase could be deduced with the help of the pressure dependence of the *c/a* ratio. At room pressure, in $CaWO_4$ and $YLiF_4$ the cation coordination is [8 – 4]. The decrease of the *c/a* ratio in CaWO$_4$ with increasing pressure leads to a structure with cation coordination [6 – 6] as it is indeed in the wolframite structure. On the contrary, the increase of the *c/a* ratio in YLiF$_4$ with increasing pressure leads to a structure with different cation coordination as it occurs in M-fergusonite, with a cation coordination between [8 – 4] and [8 – 6]. It is interesting to note that a similar increase of the *c/a* ratio with increasing pressure has been recently calculated in the ionic perrhenates AgReO$_4$ and NaReO$_4$ **[22]**. The scheelite perrhenates usually transform at high pressures to the orthorhombic pseudoscheelite structure, whose cation coordination is similar to that of scheelite and M-fergusonite structures.

Experimental results agree with these previous considerations since the *c/a* ratio for CaWO$_4$ is larger than for YLiF$_4$. This indicates that the WO$_4$ group is more covalent than the LiF$_4$ group, as it is indeed. As a consequence of this different covalence, there are smaller cation-cation repulsion forces in CaWO$_4$ than in YLiF$_4$ at atmospheric pressure. However, with increasing pressure cation-cation repulsion forces become dominant in CaWO$_4$ while packing considerations become dominant in YLiF$_4$ due to the increase in the covalence of the Y-F and Li-F bonds with increasing pressure. The decrease of the axial ratio in CaWO$_4$ upon compression, specially above 5 GPa, could be related to just a small increase of the cation-cation electrostatic repulsion, which can be tentatively ascribed to a change in the electronic density around the Ca and W atoms. Furthermore, the change of cation coordination at the scheelite-to-wolframite transition could be originated by a *s-d* charge transfer effect **[3]**:

i) at low-pressure the occupation of the *s* orbital is favored **[23]**, resulting in a more symmetrical distribution; and

ii) at high pressures the occupation of a localized *d* orbital might induce a strong distortion, which would favor the transition to the wolframite structure as it happens in the temperature-driven tetragonal-to-monoclinic transition of BiVO$_4$ **[24]**.

On the other hand, the increase of the axial ratio in YLiF$_4$ has been previously ascribed to the big ionic character of the fluorine bonds as compared to those formed by oxygen, the Y-F bond being less ionic and considerably less compressible than the Li-F bond. Therefore, the increase of the tetragonal distortion with increasing pressure was understood due to the big initial compressibility of the Li-F bond. The saturation of the increase of the axial ratio in YLiF$_4$ above 6 GPa could be related to the stiffening of the Li-F bond, which would lead to a small increase of the cation-cation repulsion at the Li sites with increasing pressure due to the small size of Li atoms.

Another interesting fact is that the reconstructive scheelite-to-wolframite transition in CaWO$_4$ occurs together with a collapse of both the W-O bonds (1.798 Å → 1.698 Å) and the Ca-O bonds (2.293 Å → 2.183 Å and 2.379 Å → 2.272 Å) at the phase transition **[3]**. The reduction observed in the Ca-O distance is coherent with the occurrence of a change of the Ca ionic radii from 1.12 Å (when Ca to O coordination is 8 in the scheelite phase) to 1 Å (when Ca to O coordination is 6 in the wolframite phase) whereas the reduction of the W-O distances could be related to a change of the character of the bond. On the other hand, the fact that both bonds collapse at the transition is reflected in the fact that the axial ratio remains nearly constant during the transition (the *2c/a* ratio of the high-pressure wolframite phase is equal to the *c/a* ratio of the scheelite phase before the transition **[3]**). In addition, the scheelite-to-wolframite transition produces a distortion of the planes perpendicular to *c*. Basically, the crystal is deformed along one direction, making $b > a$. This fact is likely related to a tilting of the W-O polyhedra that could easily explain the occurrence of the scheelite-to-wolframite transition.

**3.2 Phase transition mechanisms**

In order to understand the scheelite-to-wolframite and scheelite-to-fergusonite transitions we have to note that:

i) the ionic-covalent bonds in the $ABX_4$ fluorides are much weaker than the more covalent bonds in $ABX_4$ oxides;

ii) long bonds are usually softer and more compressible than short ones;

iii) under compression almost all bonds become shorter (and most of them stronger); and

iv) upon the application of pressure cation-cation repulsive interaction increases considerably.

On this basis, it is commonly accepted that the atomic structures of $ABX_4$ compounds under high pressures should tend to structures with a higher and equal coordination of both A and B cations **[20]**. The structural phase transitions shown by $CaWO_4$ and $YLiF_4$ point towards this direction because both high-pressure monoclinic phases (wolframite and M-fergusonite) show larger average cation coordination than the scheelite structure ([8 - 4]). In the wolframite structure, each A and B cation is in an approximately octahedral coordination surrounded by six near X sites **[3, 7]**; i.e., with cation coordination [6 - 6], as shown in **Fig. 6(a)**. A view of the cations in the wolframite structure is shown in **Fig. 6(b)**. The [6 - 6] coordination in the wolframite phase suggests similar strengths for the forces associated to the W-O and Ca-O bonds in such structure.

In addition, this fact also points towards an increase of the coordination number around W cations with increasing pressure in the scheelite phase of $CaWO_4$. On the other hand, in the M-fergusonite structure, each A cation is surrounded by eight X anions and each B cation is surrounded by four X sites and two additional near X sites. Therefore, the M-fergusonite structure is considered as a deformed scheelite structure, which can be described as an intermediate structure between [8 - 4] and [8 - 6] cation coordination. **Fig. 7** shows two views of the cation arrangement in the M-fergusonite structure.

The mechanism of the scheelite-to-wolframite transition in $CaWO_4$ around 11 GPa is of reconstructive nature and involves the destruction of both the diamond-like structures of Ca and W cations of the scheelite structure at the transition pressure. This reconstructive transition is due to the similar cation-cation repulsion forces at the Ca and W sites at the transition pressure that corresponds to similar Ca-O and W-O forces at the phase transition pressure. It is noteworthy the similarity of Ca-O and W-O forces at the transition pressure despite the ionic and the covalent characters of the Ca-O and W-O bonds at ambient pressure, respectively **[25]**. From a short-range point of view, this phase transition mechanism is related to a shift of the W cation from the center of the $WO_4$ tetrahedron towards the center of the $WO_6$ octahedron, and it is characterized by:

i) a motion of the W atoms from the center of the W-O tetrahedra along the *b* direction; and

ii) a shear displacement of its second neighbors O atoms.

**Fig. 8(a)** shows a schematic representation of the scheelite-to-wolframite transformation mechanism here proposed. **Fig 8(b)** shows the (100) projection of a section of the scheelite structure compared with that of a portion of the wolframite structure in order to better illustrate the transformation. We believe that the process leading to the scheelite-to-wolframite transition is the following: At low pressure, the weak Ca-O bonding of the $CaO_8$ polyhedra absorb much of the pressure while the $WO_4$ tetrahedra remain as rigid units. When reaching around 10 GPa, the Ca-O bond length has decreased much more than the W-O bond length so as to become as strong as the W-O bond (see **Fig. 5**). Upon further application of pressure, the W-O tetrahedral units are tilt and distorted and the [010] planes shear forming a distorted "Star of David" (see **Fig. 8(b)**). This configuration is characteristic of a cation in octahedral coordination when viewed perpendicular to the c axis of the scheelite (with four O atoms at 1.698 Å from W and two O atoms at 1.898 Å from W in the distorted octahedron).

On the other hand, the mechanism of the scheelite-to-M-fergusonite transition in $YLiF_4$ around 11 GPa is of martensitic nature and it is preceded by a reversible polytype phase transition at 6 GPa. The $LiF_4$ tetrahedra in the scheelite structure of $YLiF_4$ form an angle $\phi = 29°$ with respect to the main *a* axis at ambient pressure **[18]**. With increasing pressure, the Li-F distance decreases till the $LiF_4$ tetrahedra become rigid around 6 GPa. At higher pressures, the stiffening of the Li-F bond (see **Fig. 4** in **Ref. 5**) and the progressive decrease of the *a* lattice parameter above 6 GPa is only possible if there is a gradual rotation of the $LiF_4$ tetrahedra around the tetragonal *c* axis; i.e. in the *a-b* plane, towards larger angles. This fact means that the $LiF_4$ tetrahedra can only rotate till the maximum value of $\phi = 45°$ compatible with the scheelite structure and the reduction of the *a* lattice parameter. This rotation can be considered as a reversible phase transition from a polytype-I to a polytype-II scheelite structure. The polytype I is characterized by a setting angle $\phi = 29º$, closer to the higher symmetry zircon structure with $\phi = 0º$, while the polytype II is characterized by an angle $\phi$ ($29º < \phi < 45º$).

The reversible phase transition from scheelite polytype-I to polytype-II at 6 GPa in YLiF$_4$ is induced by polyhedral tilting (in this case rotation in the *a-b* plane). This phase transition is possible due to the softening of one of the traslational T(E$_g$) modes of the scheelite phase that involves a rotation of the LiF$_4$ tetrahedra in the plane perpendicular to the c axis **[26]**. It is worth noting that the softening of the T(E$_g$) mode of the zircon phase of YVO$_4$ is also the responsible for the zircon-to-scheelite phase transition above 7.5 GPa **[27]**, since the VO$_4$ tetrahedra in the zircon phase form an angle $\phi = 0º$ with respect to the main *a* axis while that angle is always different from $\phi = 0º$ in the scheelite structure.

A characteristic of this kind of reversible transitions is that the low-pressure structure (with higher symmetry) shows a certain degeneration of the vibrational modes, which disappears once the phase transition to the low-symmetry structure is accomplished **[28]**. A splitting of several Raman modes above 6 GPa that was initially overlapped is indeed observed **[12,13]**. Furthermore, this structural change around 6 GPa in YLiF$_4$ is reflected in a slight modification of the pressure coefficients of the frequency of some Nd$^{3+}$ crystal-field transitions above 6 GPa **[6]**.

Reversible transitions show no major change in cation coordination, except for subtle displacements in the cation coordination of those cations with larger coordination number. They also occur in a sudden and reversible way leaving the crystal lattice undamaged during the transformation and with reduced volume changes. Furthermore, the reversible transitions are usually followed by twinning; i.e., a mixture of different lattice orientations of the new crystals due to the loss of a symmetry element in the phase transition. This fact can affect the accurate determination of the lattice parameters in the new structure and could be related to the strange behavior of the distances estimated in **Ref. 5** from high-pressure X-ray diffraction measurements between 6 and 11 GPa. Moreover, the reversible phase transition around 6 GPa is coherent with the martensitic phase transition occurring at 10 GPa in YLiF$_4$ since both reversible and martensitic transitions are common in ionic compounds **[29, 30]**. As a matter of fact, they have been also found in other similar compounds like KAlF$_4$ and RbAlF$_4$ as a function of temperature and pressure **[31, 32]**.

Finally, the martensitic scheelite-to-M-fergusonite transition is a shear transformation in which the initial structure is partially conserved while certain sheets or pieces of the previous structure are slightly shifted. In YLiF$_4$, it involves a shift of long zigzag chains of B (Li) cations either along [100] or along [010] directions of the scheelite structure (see the schematic model shown in **Fig. 9**). Previous studies suggest that layer shifts along the [100] direction are energetically more favorable than shifts along [010] direction **[33]**. The large shift of B cations in YLiF$_4$, in contrast to what is observed in CaWO$_4$, is due to the mainly ionic character of the Li-F bond, which is much weaker than the covalent W-O in CaWO$_4$. This fact makes the Li atoms less tightly bound than the W atoms in the scheelite structure. This type of transitions is quick and in certain cases the crystal is undamaged despite the symmetry of the final structure is lower than that of the previous one. These transformations are usually reversible; i.e., the initial structure is recovered on release of pressure, but they usually show a certain hysteresis; a behavior indeed found in the scheelite-to-M-fergusonite phase transition in YLiF$_4$ above 10 GPa [6].

This phase transition takes place because cation-cation repulsion increases considerably above 10 GPa, especially at the Li sites, what leads to the destruction of the Li diamond-like structure at the transition pressure whereas the Y diamond-like structure is preserved and slightly distorted. This well-known high-pressure structure is related to the scheelite structure, since it can be considered as a distorted scheelite (see the comparison view of both structures in **Fig. 9**), and conversely the scheelite structure can be viewed as a tetragonal fergusonite [30]. The larger increase of the repulsion at the Li sites as compared to the Y sites is likely due to a major change in the electronic density around the Y atoms with increasing pressure. This change in the electronic density around the Y atoms occurs because of the *s-d* charge transfer previously commented [34] and does not affect to Li atoms. The slight distortion of the Y diamond-like lattice and the shift of Li cations allow us to understand the martensitic second-order phase transition nature of the scheelite-to-M-fergusonite transition that proceeds without volume change, as demonstrated by Gingerich and Bair [35].

Several additional facts support the above described mechanisms for the scheelite-to-monoclinic phase transitions in CaWO$_4$ and YLiF$_4$. There is a vision that the oxide scheelites can be considered as having a complex layer-like structure, the layers being perpendicular to the *c* axis and formed by a CsCl-type arrangement of A and BO$_4$ ions [33]. This view of oxide scheelites as complex layer structures is supported by the large values of the *c/a* ratios of these compounds at ambient conditions, as compared to those of nearly ideal fluoride scheelites. Therefore, the decrease of the axial ratio in CaWO$_4$ with increasing pressure tending to the ideal structure is in agreement with the tendency of several scheelite oxides to transform to the wolframite structure with increasing pressure [36]. In summary, oxide scheelites show a tendency towards a layer-like structure, unlike fluoride scheelites. In this sense, the high-pressure scheelite-to-wolframite transition is expected in CaWO$_4$ because the wolframite structure has also a layer-like structure, unlike the M-fergusonite one. The layer-like structure of the wolframite structure along the *a* direction can be observed in **Fig. 6(b)**.

The different tendency of the oxides and fluorides scheelites towards layer-like structure due to their different nature is also reflected in the thermal expansion coefficients of oxide and fluoride scheelites. In fluoride scheelites, the $\alpha_{11}$ tensor component of the thermal expansion is greater than the $\alpha_{33}$, whereas in oxide scheelites the contrary is true, as it is usual in layer-like crystals [11, 37]. Furthermore, we believe that the different high-pressure structures observed in both compounds are related to the different nature of bonds in CaWO$_4$ and YLiF$_4$ and exhibit a link with the different behavior of the axial compressibilities and the different soft modes observed in both compounds. In this respect, Blanchfield *et al.* noted the instability of these two compounds under application of shear stresses, as deduced from the softening of several lateral modes [10, 11]. However, the soft modes with bigger softening are not the same in CaWO$_4$ and YLiF$_4$, pointing out a significant difference between both compounds. This result agrees with recent results of ion rigid calculations in YLiF$_4$ [38] and with recent Raman scattering measurements under pressure [9, 12, 13]. In CaWO$_4$ a softening of one of the traslational zone center T(B$_g$) modes as pressure increases has been observed [9], while in YLiF$_4$ the mode that softens under pressure is one of the traslational zone center

T($E_g$) modes **[13, 38]**. These two modes are interrelated because both can be considered as external modes of the $BX_4$ tetrahedra and their frequencies are greatly affected by the substitution of the A cation **[39]**. These vibrational modes are low-frequency modes in both compounds and are associated to traslations of the $BX_4$ tetrahedra. The $B_g$ mode is associated to the vibration of the $BX_4$ tetrahedra along the tetragonal axis of the scheelite, whereas the $E_g$ mode is related to the vibration of the $BX_4$ tetrahedra in the plane perpendicular to the tetragonal axis of the scheelite. We think that the softening of these modes is indicative of the increase of the B-B distances observed in the scheelite-to-monoclinic phase transitions.

As a summary, we may conclude that the mechanisms that lead to the scheelite-to-wolframite transition are:

i) an increase of the B-B distance, due to the traslation of the $BX_4$ tetrahedra along the *c* axis but maintaining the same mass center as in the scheelite structure; and

ii) a tilt of the $BX_4$ tetrahedra respect to the *c* axis (see **Fig. 9**).

Both the increase of the B-B distance along the *c* axis of the scheelite and the tilt of the $BX_4$ tetrahedra can be associated to the softening of the $B_g$ mode.

Correspondingly, the mechanisms that lead to the scheelite-to-M-fergusonite transition are:

i) a rotation of the $BX_4$ tetrahedra around the *c* axis of the scheelite;

ii) a slight distortion of the Y diamond-like structure; and

iii) a traslation of the $BX_4$ tetrahedra along the *a* (or *b*) directions of the scheelite, leading to an increase of the B-B distance along the *b+c* (or *a+c*) direction of the scheelite (see **Figs. 1 and 9**).

Both the rotation and the traslation along the *a* or *b* direction can be associated to the softening of the $E_g$ mode.

**3.3 Size Criterion**

We have attempted to correlate the packing ratio of the anionic $BX_4$ units around the A cations and the known phase transition pressures in the scheelite $ABX_4$ compounds. **Table I** summarises the available data on pressure studies of sixteen different scheelite $ABX_4$ compounds. In **Fig. 10**, we have plotted the transition pressure vs. the $BX_4/A$ radii ratio because this ratio is the sum of the $X/A$ plus the $B/A$ effective ionic ratios. To calculate the $BX_4/A$ values (given in **Table I**), the ionic radii of A, B, and X atoms were taken from the literature **[51 - 54]**. As a result, we have observed that the phase transition pressure increases as the ratio between the ionic radii ($BX_4/A$) increases. From these data, the following equation for the transition pressure ($P_C$) as a function of the ($BX_4/A$) radii ratio can be obtained:

$$P_C \text{ (GPa)} = (1 \pm 2) + (10.5 \pm 2)(BX_4/A - 1) \qquad (1)$$

This relationship indicates that for $BX_4/A < 1$ the scheelite structure is hardly stable even at ambient pressure. To understand the physics underlying **Eq.(1)**, we have to remind that both effective ionic radii decrease in cations and anions with increasing pressure, being the radius decrease bigger for the larger anionic radii, as already commented **[19,20]**. Therefore, the $B/A$ ratio is almost constant with increasing pressure while the $X/A$ ratio decreases considerably. Consequently, it is expected that the $BX_4/A$ ratio decreases with increasing pressure and that those compounds showing a smaller $BX_4/A$ ratio should exhibit lower transition pressures. This has been already empirically found in scheelite compounds, as shown in **Table I**.

The above hypothesis for the instability of the scheelite compounds with $BX_4$/A radii ratios near or below 1 is also supported by the transition pressures found in the alkaline-earth perrhenates and periodates families **[40-42, 55]**. It has been shown that $KReO_4$, $RbReO_4$, $KIO_4$ and $RbIO_4$ crystallize in the scheelite structure. However, $TlReO_4$, $CsReO_4$, and $CsIO_4$, showing smaller $BX_4$/A ratios near 1, crystallize in a pseudoscheelite structure at ambient pressure, being this structure one of the high-pressure phases of perrhenates and periodates crystallizing in the scheelite structure at ambient pressure.

The above given observations suggest that the proposed size criterion (which effectively also applies to $A_2BX_4$ compounds **[56]**) could constitute a significant step towards unraveling the mechanisms underlying pressure-driven transformations in scheelite compounds. Particularly, this simple criterion could be useful to predict the occurrence of pressure-driven instabilities in additional scheelite compounds like, e.g $ZrGeO_4$, $NaReO_4$, and $KRuO_4$, for which **Eq. (1)** predicts the occurrence of pressure-driven phase transitions at 14.8 GPa, 11.5 GPa and 7.3 GPa, respectively. **Eq. (1)** could be also helpful to estimate pressure-driven instabilities in metastable scheelite compounds. Some of these compounds can be quenched at ambient pressure after a pressure cycle; as it occurs with $YVO_4$ **[57]**. A phase transition near 11 GPa is estimated for this compound which could be related to the zircon to scheelite phase transition observed around 8 GPa.

As regards further high-pressure phase transitions in $CaWO_4$ and $YLiF_4$, the wolframite structure of $CaWO_4$ leads to an amorphous phase above 40 GPa **[3]**. However, the M-fergusonite structure of $YLiF_4$ seems to lead to a new high-pressure phase still not determined above 17 GPa **[5, 13]**. Several reports indicate that M- to M'-fergusonite phase transitions are common in ferroelastic materials under decrease of temperature or increase in pressure **[58, 59]**, the M' phase being isostructural to the baddeleyite structure (SG: *P21/c*, No. 14, Z=2) **[1, 2]** or to the wolframite structure **[59 - 61]**. The M- to M'-fergusonite phase transition is of reconstructive-type with the unit cell of the baddeleyite structure similar to the wolframite unit cell. In this sense, the *b* axis of the M-fergusonite is almost twice that of the wolframite or the baddeleyite being the *a* and *c* axis of the M-fergusonite slightly smaller than those of the wolframite and the baddeleyite. Therefore, a phase transition to a baddeleyite (or wolframite) structure is likely expected for $YLiF_4$ above 17 GPa. The transition to the baddeleyite structure would be consistent with the baddeleyite structure shown by $MnLiF_4$ at ambient pressure, being the $Mn^{3+}$ ionic radius smaller than that of $Y^{3+}$. Thus, the decrease of the Y ionic radius with increasing pressure could increase the instability of the M-fergusonite structure leading to the baddeleyite structure. On the other hand, the transition to the wolframite structure would be also possible and it has been predicted by recent electronic structure *ab initio* calculations performed using the VASP code **[62]**. New high-pressure x-ray diffraction studies of $YLiF_4$ are required to answer definitively the new high-pressure structure.

## 4. CONCLUDING REMARKS

We report the pressure dependence of the lattice parameters and bond distances of the scheelite phase of $CaWO_4$ and compare them to those previously reported for $YLiF_4$. The comparison of the thermal expansion coefficients and the pressure coefficients found for the lattice parameters, bond distances, and Raman modes in both compounds has allowed us to understand why these two scheelites do not show the same high-pressure phase transitions. A mechanism for each of the two scheelite-to-monoclinic (wolframite or M-fergusonite) phase transitions has been proposed. Furthermore, from a comparative analysis of sixteen different scheelite compounds a close relationship between the phase transition pressures in scheelites and the $BX_4/A$ radii ratio has been found. This simple criterion can be applicable to the search of new pressure-induced transformations in scheelite compounds.

**Acknowledgements**

The authors gratefully acknowledge the contribution of A. Vegas and A. Segura, who reviewed this paper making valuable comments. We also thank Dr. J. Hu of beam line X-17C at NSLS for valuable technical advice and assistance. This work was supported by the NSF, the DOE, and the W. M. Keck Foundation. F.J.M. acknowledges financial support from the European Union under Contract No. HPMF-CT-1999-00074. D. E. also acknowledges the financial support from the MCYT of Spain through the "Ramón y Cajal" program for young scientists.

**Table I.** *Phase transition pressures and BX$_4$/A ratios for some scheelite compounds.*

| Compound | BX$_4$/A ratio | P$_C$ (GPa) | Reference |
|:---:|:---:|:---:|:---:|
| KIO$_4$ | 1.39 | 6.5 | 40 |
| RbIO$_4$ | 1.25 | 5.3 | 41 |
| AgReO$_4$ | 1.9 | 13 ± 1 | 21 |
| KReO$_4$ | 1.45 | 7.5 | 42 |
| RbReO$_4$ | 1.30 | 1.6 | 42 |
| CaWO$_4$ | 1.89 | 11 ± 1 | 3, 4 |
| SrWO$_4$ | 1.76 | 10.5 ± 2 | 2, 43, 44 |
| EuWO$_4$ | 1.76 | 8 ± 1 | 45 |
| PbWO$_4$ | 1.66 | 4.5 | 46 |
| BaWO$_4$ | 1.47 | 6.5 ± 0.3 | 43 |
| CdMoO$_4$ | 2.03 | 12 | 47 |
| CaMoO$_4$ | 1.88 | 8.2 ± 0.4 | 9 |
| SrMoO$_4$ | 1.74 | 12.5 ± 0.5 | 48 |
| PbMoO$_4$ | 1.64 | 6.5 ± 3 | 46, 49 |
| CaZnF$_4$ | 1.97 | 10 | 50 |
| YLiF$_4$ | 2.11 | 11 ± 1 | 5, 6, 13 |

**Figure captions**

**Figure 1.** Unit cell of the scheelite structure of $ABX_4$ compounds with the *a*, *b* and *c* axis. Big atoms refer to A cation (Ca, Y), medium-size atoms correspond to B cations (W, Li) and small atoms to the X anion (O, F). Numbers 1 and 2 correspond to B-B distances of the diamond-like structure along *b* + *c* and *a* + *c* directions, respectively. The $AX_8$ polyhedra and the $BX_4$ tetrahedra are shown.

**Figure 2:** EDXD pattern of the scheelite phase of $CaWO_4$ at 2 GPa. The background was subtracted. The stars mark the position of the diffraction lines of the Au pressure marker. The last line represents the difference between the measured data and the refined profile. The bars indicate the calculated positions of the $CaWO_4$ reflections.

**Figure 3.** Pressure dependence of the unit cell parameters of the scheelite structure in $CaWO_4$ and $YLiF_4$. Data for $YLiF_4$ (○) are taken from Ref. 5 and data for $CaWO_4$ are from the present study (●) and Ref. 8 (■).

**Figure 4.** Pressure dependence of the *c/a* ratio of the scheelite structure in $CaWO_4$ and $YLiF_4$. Data for $YLiF_4$ (○) are taken from Ref. 5 and Data for $CaWO_4$ are from the present study (●), Ref. 7 (♦), and Ref 8 (■). Lines are just guide for the eye.

**Figure 5.** Pressure dependence of the interatomic bonds in the scheelite structure of $CaWO_4$. The solid lines show the pressure dependence of the Ca-O bonds and the dashed lines the pressure dependence of the W-O bonds. Solid symbols are from the present study and empty symbols from Ref. 8.

**Figure 6.** (a) Wolframite structure of $CaWO_4$ with its unit cell and the *a*, *b* and *c* axis. (b) Wolframite structure of $CaWO_4$ in the a-b plane. Big black atoms refer to A cation (Ca),

grey medium-size atoms correspond to B cation (W) and small atoms to the X anion (O). The $AX_6$ octahedra, the $BX_6$ octahedra, and the shorter zig-zag cation-cation distances are also shown in (a) while anion atoms are not shown for the sake of clarity in (b). The shorter metal-metal distances are also shown in both schemes.

**Figure 7.** Schematic views of the cationic arrangement in the M-fergusonite structure. Black atoms correspond to the A cation (Y), grey atoms correspond to the B cation (Li). Anion atoms (F) are not shown for the sake of clarity. The shorter metal-metal distances are also shown.

**Figure 8.** (a) Schematic representation of the scheelite-to-wolframite model transition mechanism. (b) The (100) projection of a section of the scheelite structure compared to that of a portion of the wolframite structure. 1, 2, and 3 represent oxygens at (1/4,0,0) and 4, 5, and 6 oxygens at (-1/4,0,0).

**Figure 9.** Detail of the scheelite structure (left) in the *a-c* plane with A (Y) and B (Li) cations located in alternated planes along the *b* axis (perpendicular to the paper). Detail of the M-fergusonite structure (right) in the *c-b* plane with A and B cations located in alternated planes along *a* axis (perpendicular to the paper). The M-fergusonite structure derives from the scheelite structure when B cations shift along *a* axis of the scheelite.

**Figure 10.** Phase transition pressure in several scheelites as a function of the $BX_4$/A ratio. Symbols correspond to the data summarized in **Table I**, the solid line corresponds to the relation given in **Eq. (1)**, and the dashed lines are its lower and higher deviations.

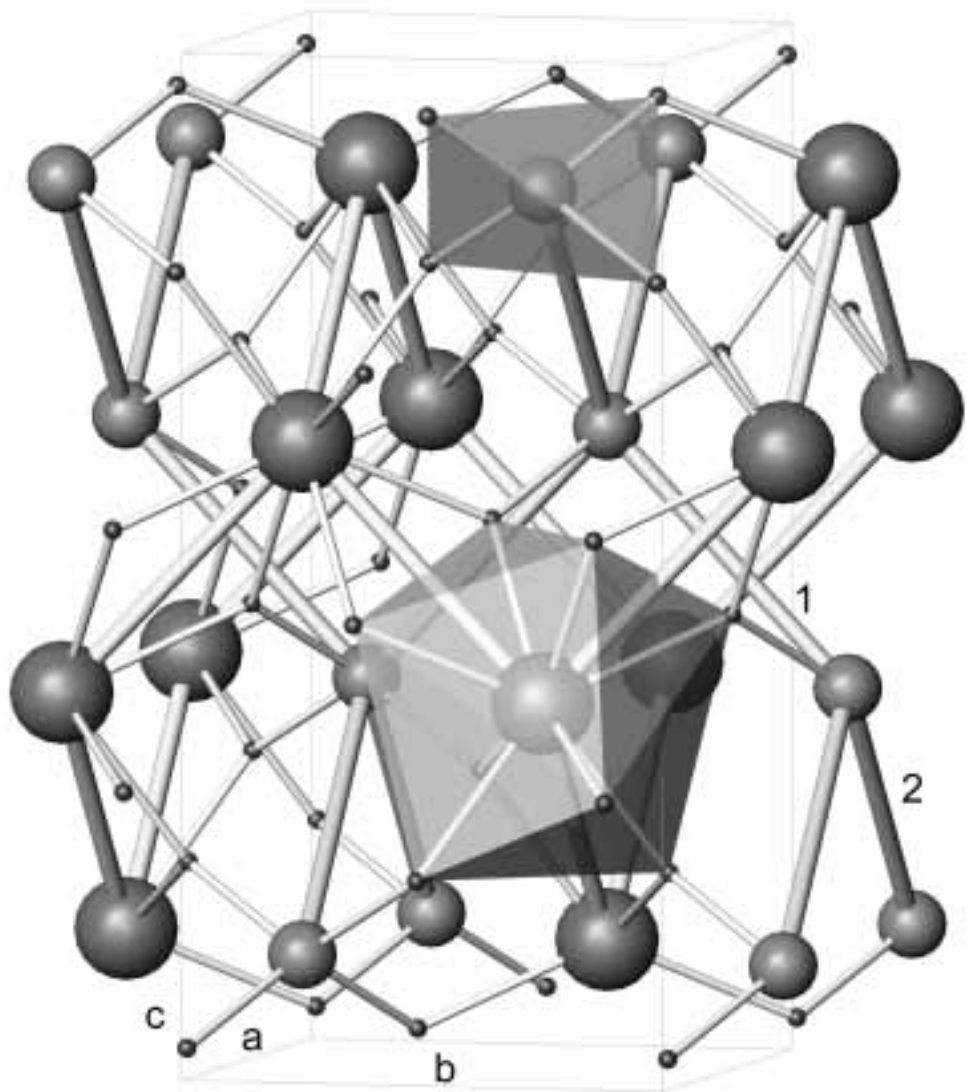

Figure 1. D. Errandonea et al.

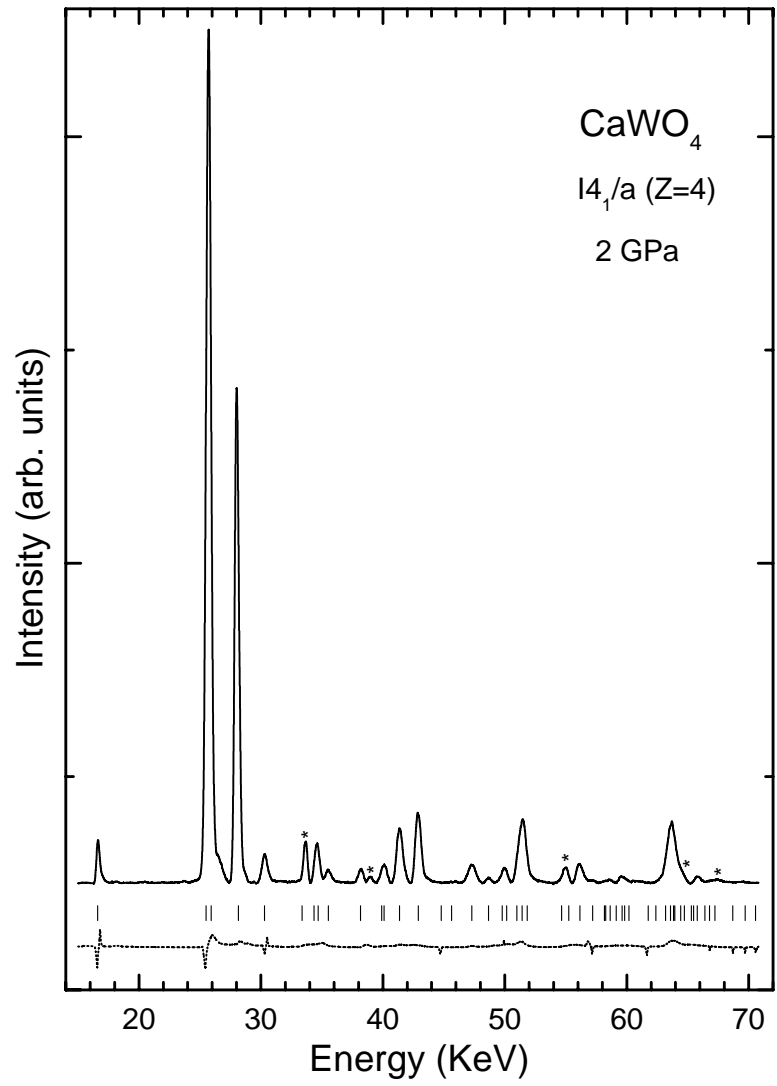

Figure 2. D. Errandonea et al.

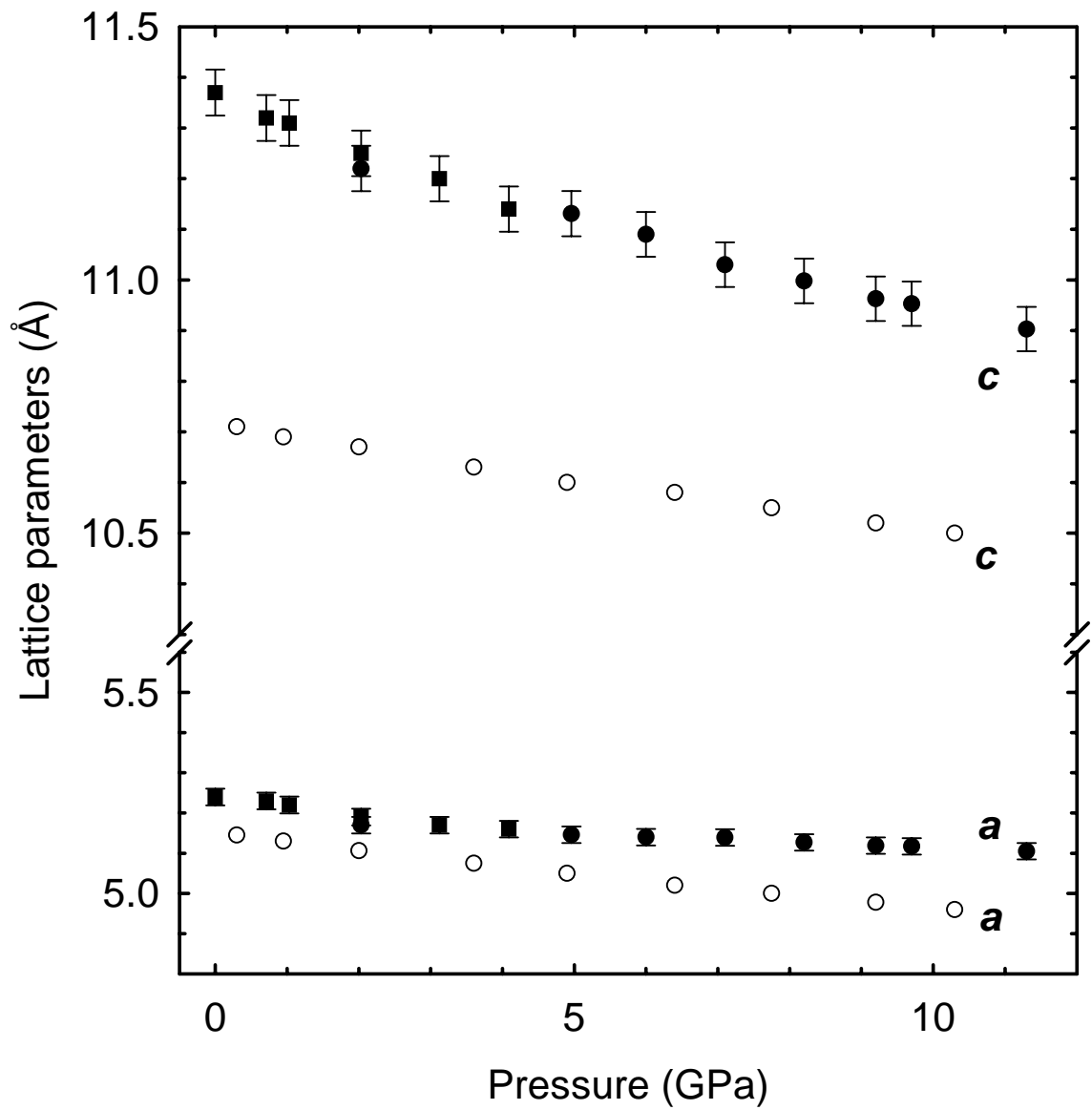

Figure 3. D. Errandonea et al.

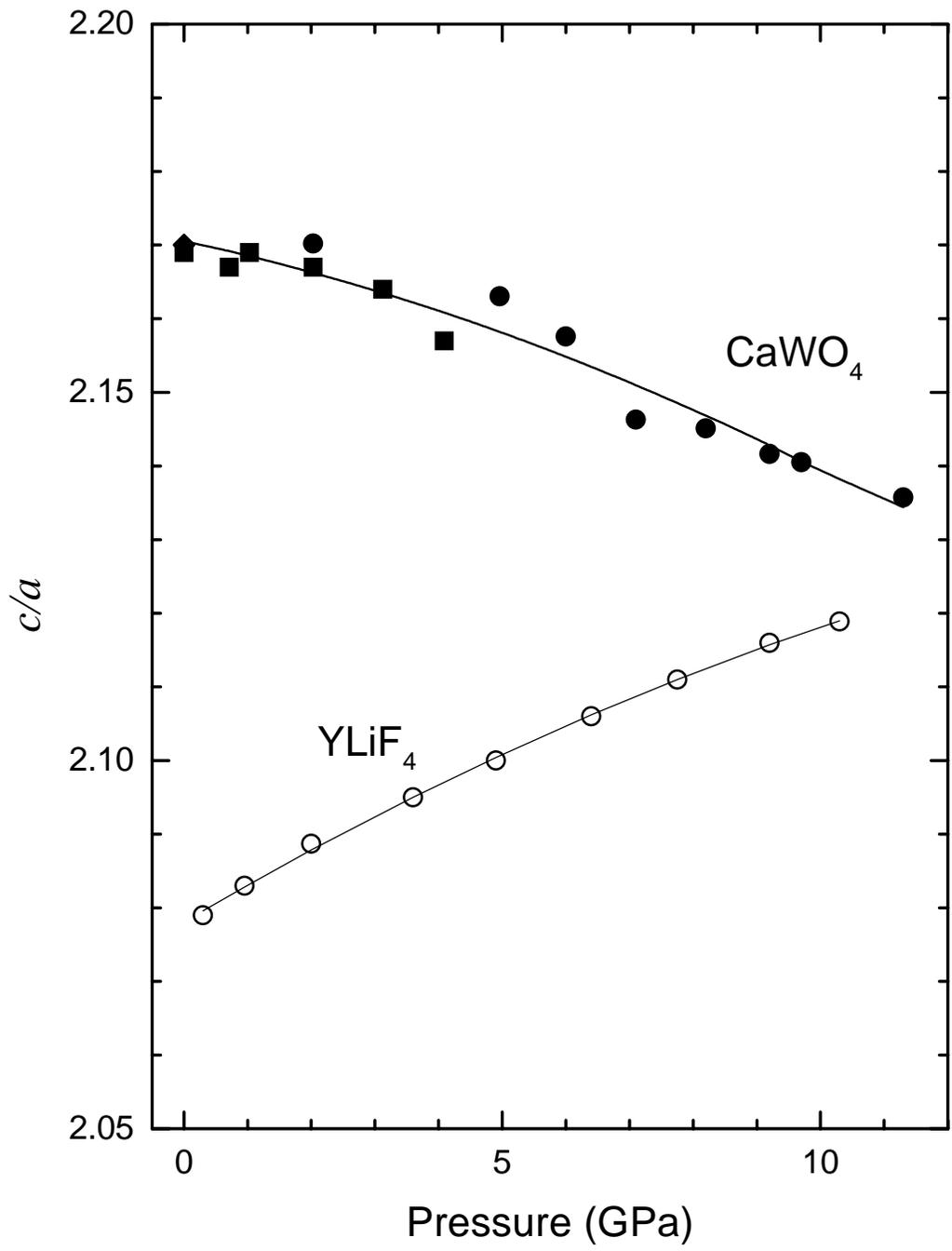

Figure 4. D. Errandonea et al.

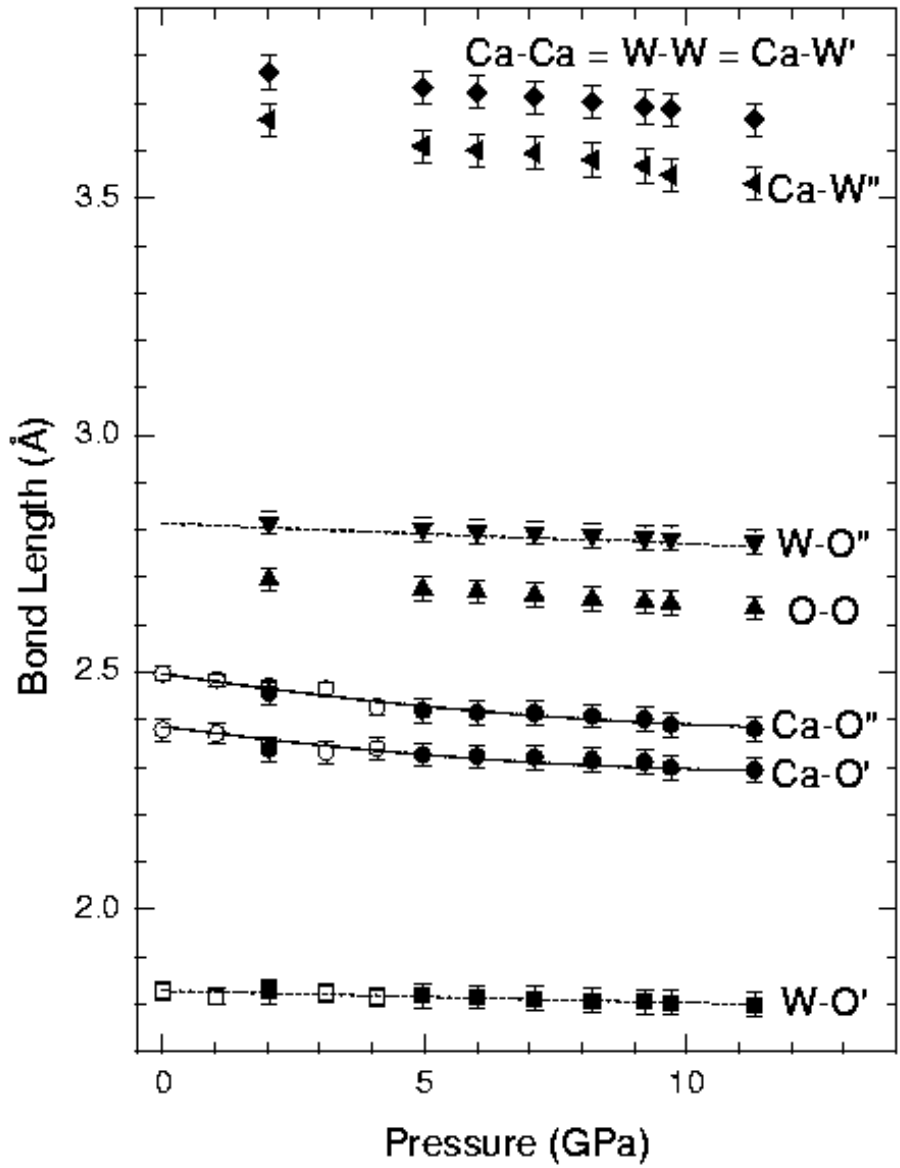

Figure 5. D. Errandonea et al.

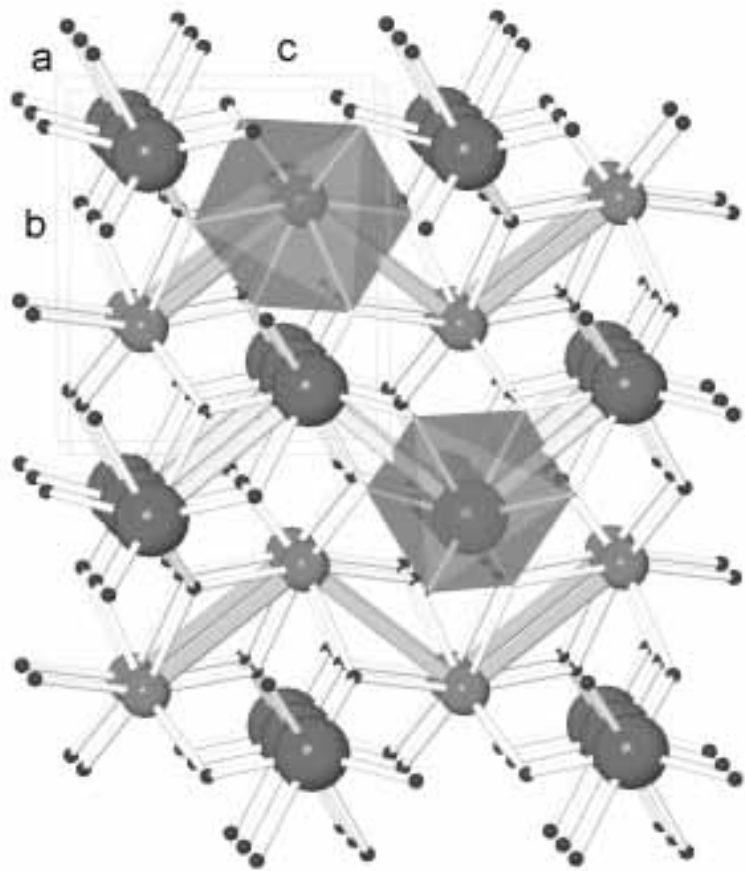

Figure 6 (a). D. Errandonea et al.

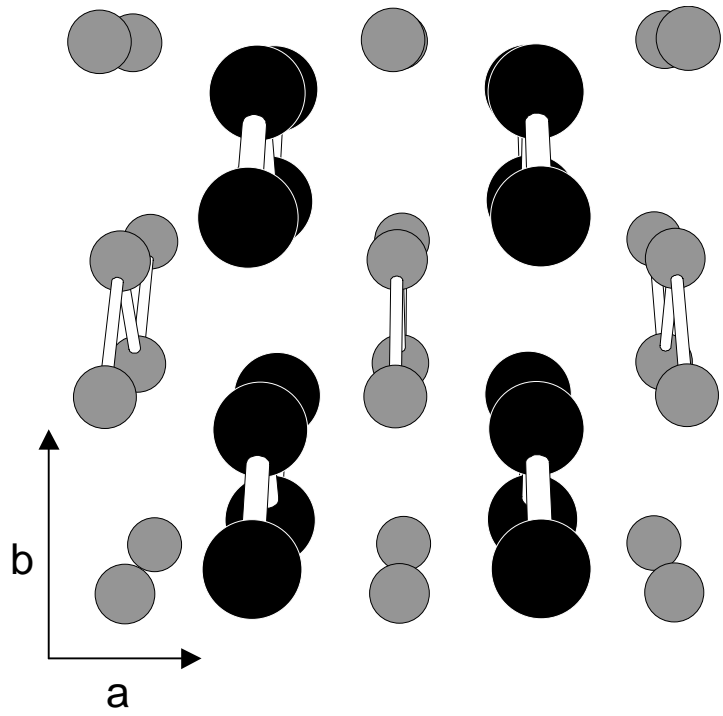

Figure 6 (b). D. Errandonea et al.

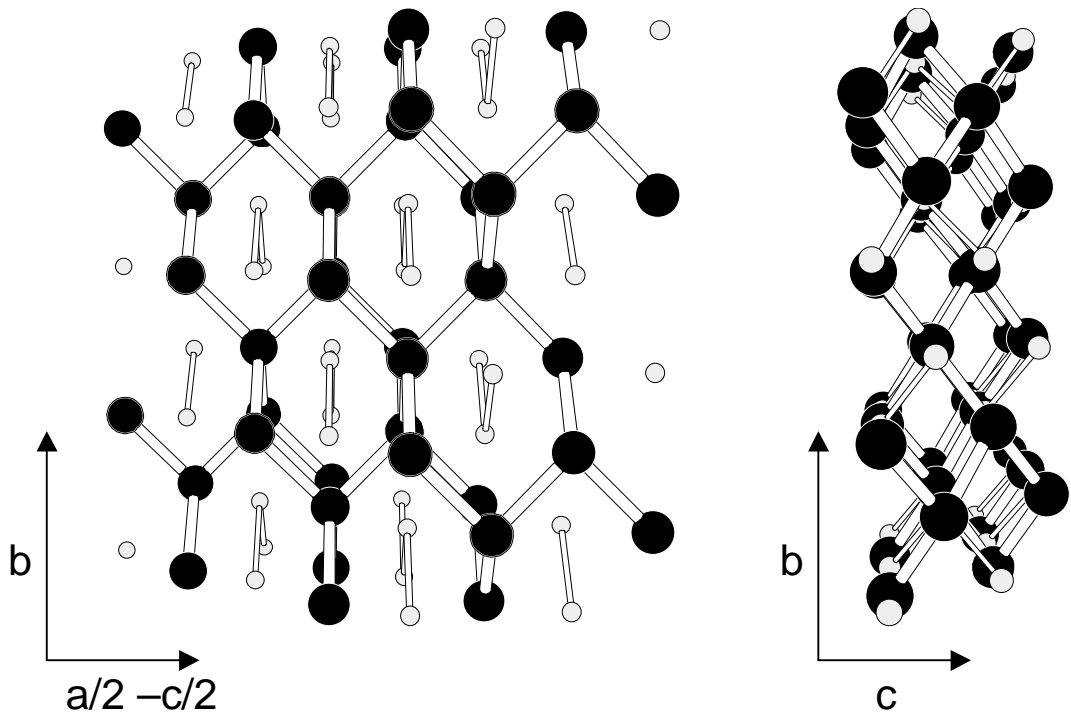

Figure 7. D. Errandonea et al.

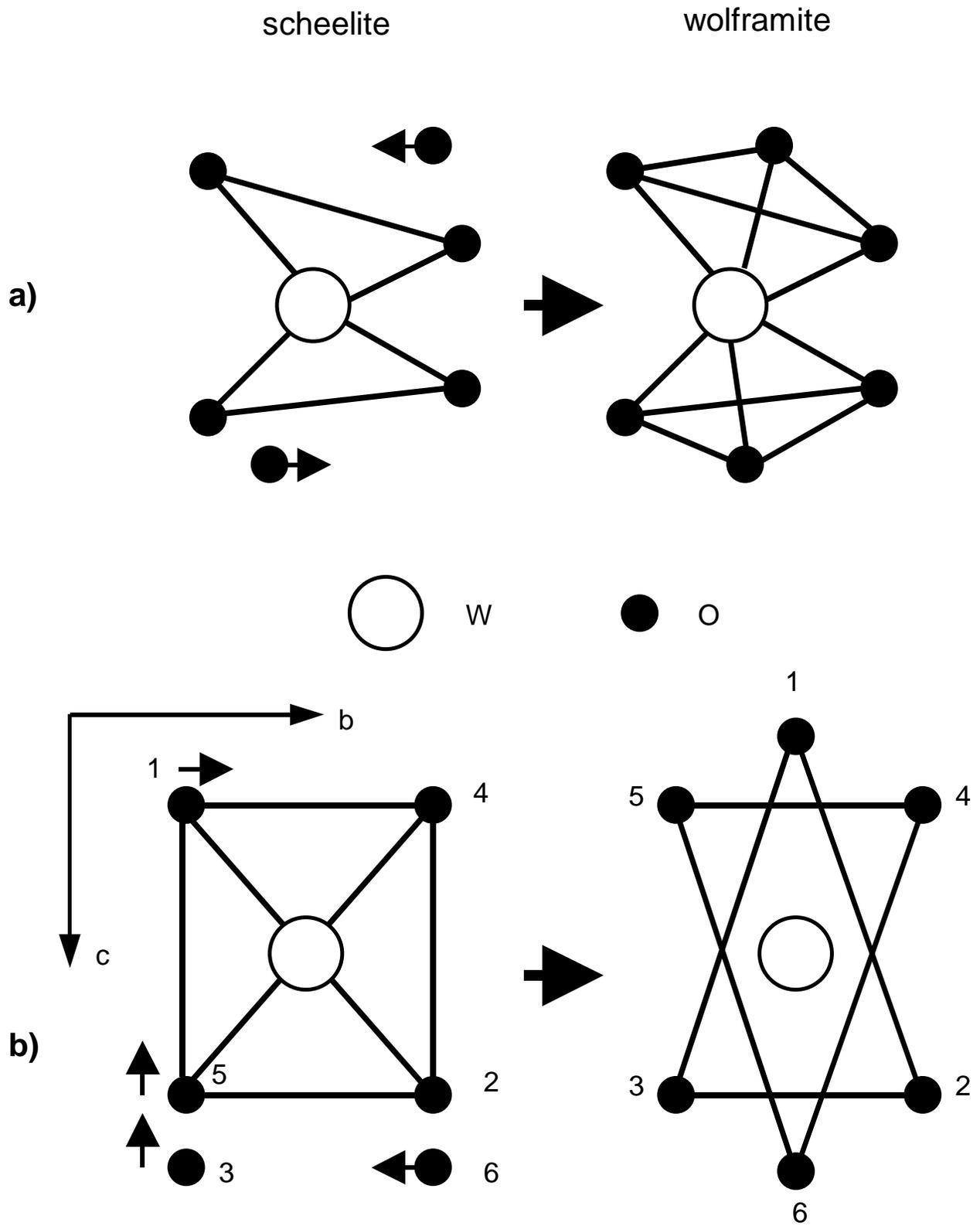

Figure 8. D. Errandonea et al.

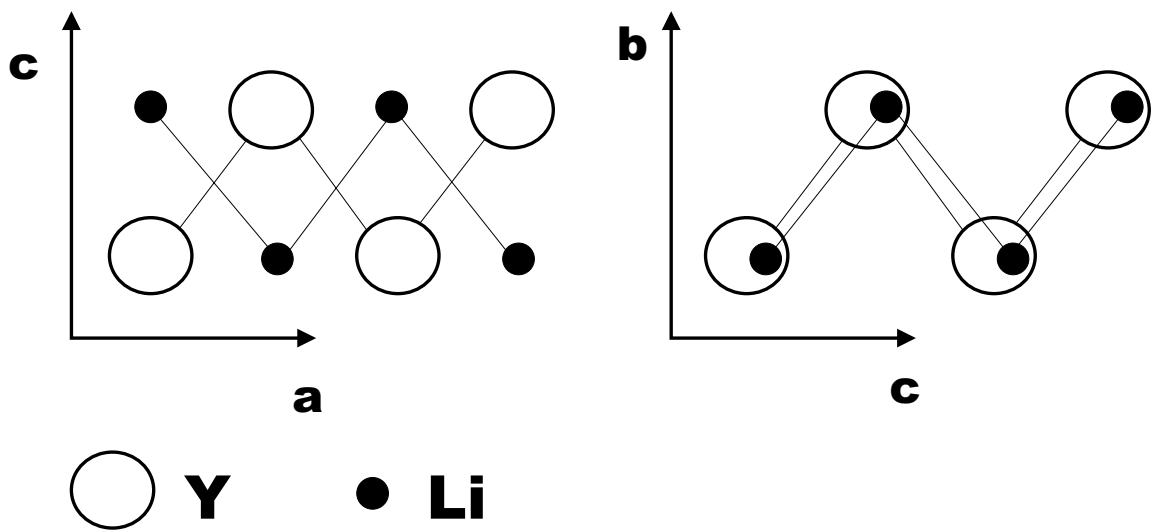

Figure 9. D. Errandonea et al.

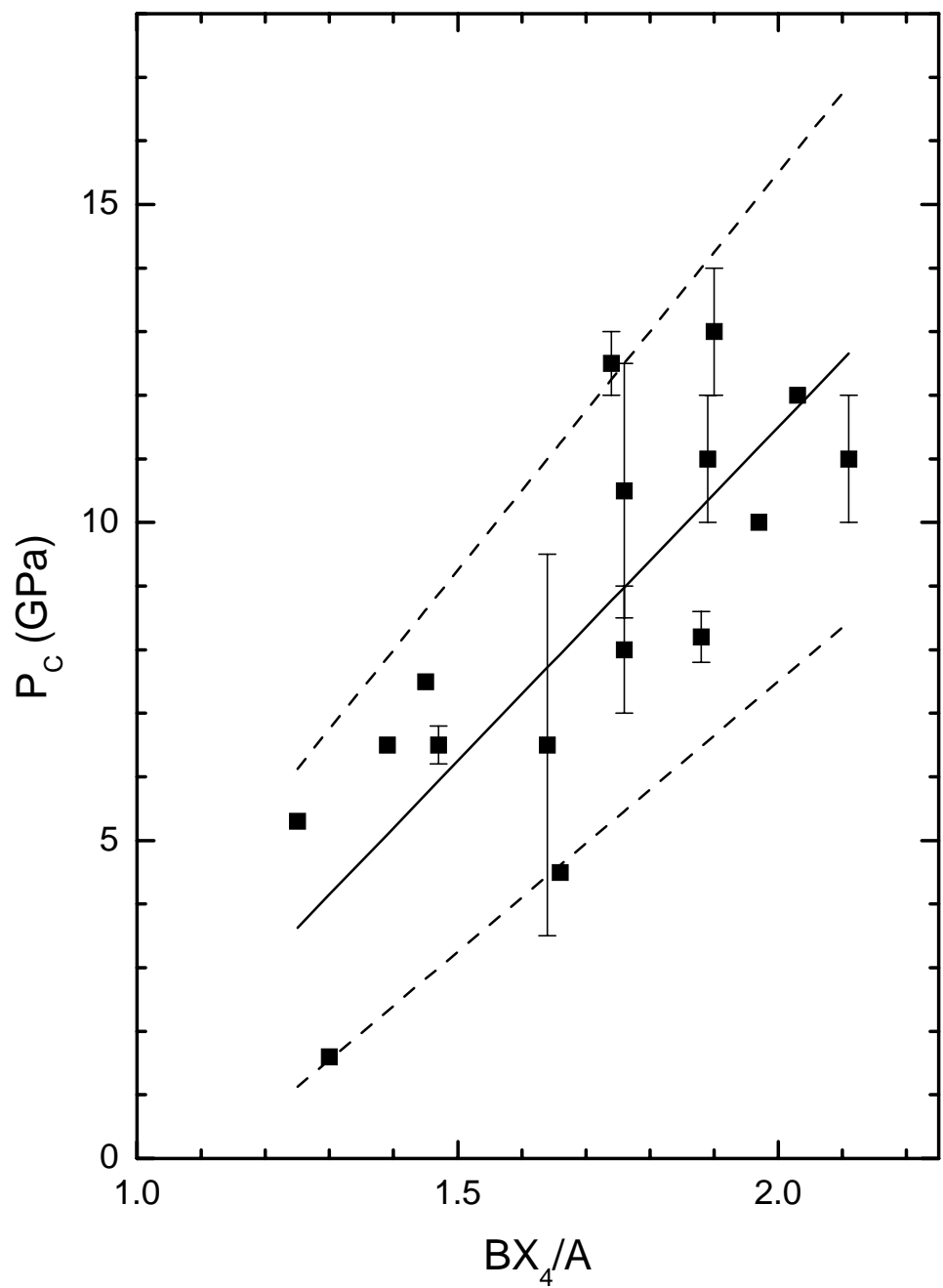

Figure 10. D. Errandonea et al.